\documentclass[preprint,prl,showpacs,groupedaddress,10pt,twocolumn]{revtex4}
\usepackage{amsfonts}
\usepackage{amsmath}
\usepackage{amssymb}
\usepackage{graphicx}
\usepackage{dcolumn}
\usepackage{bm}
\usepackage{graphics}
\usepackage{psfig}

\begin{document}

\preprint{}
\title[]{A unified approach to realize universal quantum gates in a coupled
two-qubit system with fixed always-on coupling}
\author{Zhongyuan Zhou$^{1,2}$}
\author{Shih-I Chu$^{1}$}
\author{Siyuan Han$^{2}$}
\affiliation{$^{1}$Department of Chemistry, University of Kansas, Lawrence, KS 66045\\
$^{2}$Department of Physics and Astronomy, University of Kansas, Lawrence,
KS 66045}

\begin{abstract}
We demonstrate that in a coupled two-qubit system any single-qubit gate can
be decomposed into two \textit{conditional} two-qubit gates and that any 
\textit{conditional} two-qubit gate can be implemented by a manipulation
analogous to that used for a controlled two-qubit gate. Based on this we
present a unified approach to implement universal single-qubit and two-qubit
gates in a coupled two-qubit system with fixed always-on coupling. This
approach requires neither supplementary circuit or additional physical
qubits to control the coupling nor extra hardware to adjust the energy level
structure. The feasibility of this approach is demonstrated by numerical
simulation of single-qubit gates and creation of two-qubit Bell states in
rf-driven inductively coupled two SQUID flux qubits with realistic device
parameters and constant always-on coupling.
\end{abstract}

\received[Received: ]{June 29, 2005}
\maketitle

During the past decade, a variety of physical qubits have been explored for
possible implementation of quantum gates. Of those solid-state qubits based
on superconducting devices have attracted much attention because of their
advantages of large-scale integration and easy connection to conventional
electronic circuits \cite%
{Leggett02,Clarke03,Blatter03,Nakamura1999,Vion2002,Yu2002,Han2001,Martinis2002,Friedman2000,Wal2000,Chiorescu2003,Pashkin2003,Yamamoto2003,Berkley2003,McDermott05}%
. Superconducting single-qubit gates \cite%
{Nakamura1999,Vion2002,Yu2002,Han2001,Martinis2002,Friedman2000,Wal2000,Chiorescu2003}
and two-qubit gates \cite{Pashkin2003,Yamamoto2003,Berkley2003,McDermott05}
have been demonstrated recently.

However, building a practical quantum computer requires to operate a large
number of multi-qubit gates simultaneously in a coupled multi-qubit system.
It has been demonstrated that any type of multi-qubit gate can be decomposed
into a set of universal single-qubit gates and a two-qubit gate, such as the
controlled-NOT (CNOT) gate \cite{Nielsen2000,Barenco1995}. Thus it is
imperative to implement the universal single-qubit and two-qubit gates in a
multi-qubit system with the minimum resource and maximum efficiency \cite%
{Plourde04}.

Implementing universal single-qubit gates and two-qubit gates in coupled
multi-qubit systems can be achieved by turning off and on the coupling
between qubits \cite{Plourde04,Averin03,Blais03}. In these schemes,
supplementary circuits were required to control inter-qubit coupling.
However, rapid switching of the coupling results in two serious problems.
The first one is gate errors caused by population propagation between
qubits. Because the computational states of the single-qubit gates are\ not
a subset of the eigenstates of the two-qubit gates the populations of the
computational states propagate from one qubit to another when the coupling
is changed, resulting in additional gate errors. The second one is
additional decoherence introduced by the supplementary circuits \cite%
{ZhouXingxiang2002,Benjamin2002}. This is one of the biggest obstacles for
quantum computing with solid-state qubits, particularly in coupled
multi-qubit systems \cite{Clarke03}. In addition, the use of supplementary
circuits also significantly increase the complexity of fabrication and
manipulation of the coupled qubits.

To circumvent these problems, a couple of alternative schemes, such as those
with untunable coupling \cite{ZhouXingxiang2002} and always-on interaction 
\cite{Benjamin2003}, have been proposed. In the first scheme, each logic
qubit is encoded by extra physical qubits and coupling between the encoded
qubits is constant but can be turned off and on effectively by putting the
qubits in and driving them out of the interaction free subspace. In the
second scheme, the coupling is always on but the transition energies of the
qubits are tuned individually or collectively. These schemes can overcome
the problem of undesired population propagation but still suffer from those
caused by the supplementary circuits needed to move the encoded qubits and
tune the transition energies. Moreover, the use of encoded qubits also
requires a significant number of additional physical qubits.

In this Letter, we present a unified approach to implement universal
single-qubit and two-qubit gates in a coupled two-qubit system with fixed
always-on coupling. In this approach, each single-qubit gate is realized via
two \textit{conditional} two-qubit gates and each \textit{conditional}
two-qubit gate is implemented with a manipulation analogous to that used for
a controlled two-qubit gate (e.g., CNOT) in the same subspace of the coupled
two-qubit system without additional circuits or physical qubits. Since the
computational states of the single-qubit gates are a subset of those of the
two-qubit gates the gate errors due to population propagation are completely
eliminated. The effectiveness of the approach is demonstrated by numerically
simulating the single-qubit gates and creating the Bell states in a unit of
inductively coupled two superconducting quantum interference device (SQUID)
flux qubits with realistic device parameters and constant always on coupling.

Consider a basic unit consisting of two coupled qubits which we call a
control qubit ($\mathcal{C}q$) and a target qubit ($\mathcal{T}q$) for
convenience. An eigenstate of the coupled qubits is denoted by $%
|n)=\left\vert ij\right\rangle $, which can be well approximated by the
product of an eigenstate of $\mathcal{C}q$, $\left\vert i\right\rangle $,
and that of $\mathcal{T}q$, $\left\vert j\right\rangle $, $|n)\equiv
\left\vert ij\right\rangle =\left\vert i\right\rangle \left\vert
j\right\rangle $, for weak inter-qubit coupling. The computational states of
the coupled qubits are $|1)=\left\vert 00\right\rangle $, $|2)=\left\vert
01\right\rangle $, $|3)=\left\vert 10\right\rangle $, and $|4)=\left\vert
11\right\rangle $, which correspond to the eigenstates for $i,$ $j=0,$ $1$
and $n=1,$ $2,$ $3,$ $4$, respectively. In general, the result of an
operation on $\mathcal{C}q$ depends on the state of $\mathcal{T}q$ since $%
\mathcal{C}q$ is coupled to and hence influenced by $\mathcal{T}q$. Suppose $%
U$ is a unitary operator acting on a single qubit in the four-dimensional
(4D) Hilbert space spanned by $|n)$. In order to perform $U$ on $\mathcal{C}%
q $ one wants the state of $\mathcal{C}q$ evolves according to $U$
independent of the state of $\mathcal{T}q$ which is left unchanged. This
operation is denoted by $U^{\left( 2\right) }=U_{C}^{\left( 1\right)
}\otimes I_{T}^{\left( 1\right) }$, where, the 4$\times $4 unitary matrix $%
U^{\left( 2\right) }$ and the 2$\times $2 unitary matrix $U_{C}^{(1)}$ are
the representations of $U$ in the Hilbert space of the coupled qubits and in
the subspace of $\mathcal{C}q$ while the 2$\times $2 unitary matrix $%
I_{T}^{\left( 1\right) }$ is the representation of the identity operation in
the subspace of $\mathcal{T}q$. If the matrix elements of $U_{C}^{\left(
1\right) }$ are $u_{ij}$ ($i,j=1,2$) the operation can be decomposed into
two operations as%
\begin{equation}
U^{\left( 2\right) }=\left( 
\begin{array}{cccc}
1 & 0 & 0 & 0 \\ 
0 & u_{11} & 0 & u_{12} \\ 
0 & 0 & 1 & 0 \\ 
0 & u_{21} & 0 & u_{22}%
\end{array}%
\right) \left( 
\begin{array}{cccc}
u_{11} & 0 & u_{12} & 0 \\ 
0 & 1 & 0 & 0 \\ 
u_{21} & 0 & u_{22} & 0 \\ 
0 & 0 & 0 & 1%
\end{array}%
\right) =U_{1}^{\left( 2\right) }U_{0}^{\left( 2\right) },  \label{x2}
\end{equation}%
where, the 4$\times $4 unitary matrices $U_{0}^{\left( 2\right) }$ and $%
U_{1}^{\left( 2\right) }$ are also in the 4D Hilbert space of the coupled
qubits. Since $U_{0}^{\left( 2\right) }$ only involves the states $%
\left\vert 00\right\rangle $ and $\left\vert 10\right\rangle $ and $%
U_{1}^{\left( 2\right) }$ only involves $\left\vert 01\right\rangle $ and $%
\left\vert 11\right\rangle $ they represent operations on $\mathcal{C}q$
when $\mathcal{T}q$ is in the state $\left\vert 0\right\rangle $ and $%
\left\vert 1\right\rangle $, respectively. If the state of $\mathcal{T}q$ is 
$\left\vert 0\right\rangle $ ($\left\vert 1\right\rangle $), $U_{0}^{\left(
2\right) }$ ($U_{1}^{\left( 2\right) }$) implements the desired operation $U$
while $U_{1}^{\left( 2\right) }$ ($U_{0}^{\left( 2\right) }$) does nothing.
Thus $U_{0}^{\left( 2\right) }$ and $U_{1}^{\left( 2\right) }$ represent two
conditional two-qubit gates of the coupled qubits. Eq.(\ref{x2}) shows that
for the coupled qubits \textit{any single-qubit gate can be realized via two
conditional two-qubit gates}. This property is universal and is shown
schematically in Fig. 1, where the single-qubit operation $U^{\left(
2\right) }$ and its equivalent operation via two conditional two-qubit gates 
$U_{0}^{\left( 2\right) }$ and $U_{1}^{\left( 2\right) }$ are illustrated.
The conditional gates are denoted by the open and closed circles for the
state of $\mathcal{T}q$ being $\left\vert 0\right\rangle $ and $\left\vert
1\right\rangle $, respectively \cite{Nielsen2000}.

\begin{figure}[ptb]
\begin{center}
\psfig{file=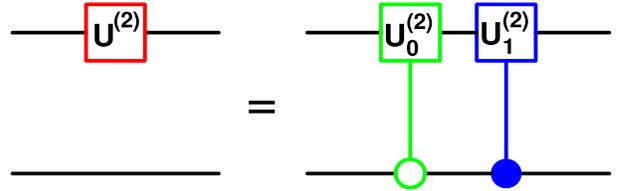,height=2.5cm,width=8cm,angle=90} 
\end{center}
\caption{Single-qubit gate $U^{(2)}$ and its equivalent operation by two
conditional two-qubit gates $U_{0}^{\left( 2\right) }$ and $U_{1}^{\left(
2\right) }$ in a coupled two-qubit system.}
\end{figure}

The most obvious advantage of implementing single-qubit gates this way is
that the conditional two-qubit gates can be realized using essentially the
same set of operations used for controlled two-qubit gates. To show this,
let us consider a rf-driven coupled two-qubit system with sufficiently
different energy-level spacings $\Delta E_{13}$, $\Delta E_{24}$, $\Delta
E_{12}$, and $\Delta E_{34}$, where $\Delta E_{nn^{\prime }}$ is the level
spacing between the states $|n)$ and $|n^{\prime })$. To implement, for
example, a single-qubit NOT gate, we apply two $\pi $ pulses to the $%
\mathcal{C}q$: the first one is resonant with $\Delta E_{13}$ and the second
with $\Delta E_{24}$. Because both pulses are largely detuned from $\Delta
E_{12}$ and $\Delta E_{34}$, they do not induce unintended population
transfer when the fields are sufficiently weak. If the state of $\mathcal{T}%
q $ is $\left\vert 0\right\rangle $ $\left( \left\vert 1\right\rangle
\right) $, the first (second) pulse accomplishes the desired transformation.
Therefore, when the two microwave pulses are applied to $\mathcal{C}q$,
either sequentially or simultaneously, the NOT gate is accomplished and the
gate time is essentially the same as that of a stand-alone $\mathcal{C}q$ if
the both pulses are applied simultaneously.

To implement a two-qubit gate such as CNOT in coupled qubits, we apply a $%
\pi $ pulse resonant with $\Delta E_{34}$ to $\mathcal{T}q$ \cite%
{zhou-asc2004}. In this case, the state of $\mathcal{T}q$ flips if and only
if the state of $\mathcal{C}q$ is $\left\vert 1\right\rangle $. Since the
energy level structure for the conditional two-qubit gates and the CNOT gate
could be the same the universal single-qubit gates and CNOT gate can be
implemented in the same fashion without adjusting inter-qubit coupling as
long as $\mathcal{C}q$ and $\mathcal{T}q$ can be addressed individually by
microwave pulses, which is rather straightforward to realize with
solid-state qubits in general and with rf SQUID flux qubits in particular.

For concreteness, we demonstrate how to realize the single-qubit and
two-qubit gates in rf-driven two SQUID flux qubits with constant always-on
coupling. The coupled flux qubits comprise two SQUIDs coupled inductively
through their mutual inductance $M$. Each SQUID consists of a
superconducting loop of inductance $L$ interrupted by a Josephson tunnel
junction characterized by its critical current $I_{c}$ and shunt capacitance 
$C$ \cite{Danilov1983}. A flux-biased SQUID with total magnetic flux $\Phi $
enclosed in the loop is analogous to a \textquotedblleft
flux\textquotedblright\ particle of mass $m=C\Phi _{0}^{2}$ moving in a
one-dimensional potential, where $\Phi _{0}$ $=h/2e$ is the flux quantum.
For simplicity, we assume that the two SQUIDs are identical. In this case, $%
C_{i}=C$, $L_{i}=L$, and $I_{ci}=I_{c}$ for $i=1,2$. The Hamiltonian of the
coupled qubits is \cite{Mooij1999,zhou-asc2004} $H\left( x_{1},x_{2}\right)
=H_{0}\left( x_{1}\right) +H_{0}\left( x_{2}\right) +H_{12}\left(
x_{1},x_{2}\right) $, where $H_{0}\left( x_{i}\right) $ is Hamiltonian of
the $i$th single qubit given by $H_{0}\left( x_{i}\right)
=p_{i}^{2}/2m+m\omega _{LC}^{2}\left( x_{i}-x_{ei}\right) ^{2}/2-E_{J}\cos
\left( 2\pi x_{i}\right) $ and $H_{12}$ is the interaction between the
SQUIDs given by $H_{12}=m\omega _{LC}^{2}\kappa \left( x_{1}-x_{e1}\right)
\left( x_{2}-x_{e2}\right) /2$. Here, $x_{i}=\Phi _{i}/\Phi _{0}$ is the
canonical coordinate of the $i$th \textquotedblleft flux\textquotedblright\
particle and $p_{i}=-i\hbar \partial /\partial x_{i}$ is the canonical
momentum conjugate to $x_{i}$, $x_{ei}=\Phi _{ei}/\Phi _{0}$ is the
normalized external flux bias of the $i$th qubit, $E_{J}=m\omega
_{LC}^{2}\beta _{L}/4\pi ^{2}$ is the Josephson coupling energy, $\beta
_{L}=2\pi LI_{c}/\Phi _{0}$ is the potential shape parameter, $\omega
_{LC}=1/\sqrt{LC}$ is the characteristic frequency of the SQUID, and $\kappa
=2M/L$ is the coupling strength. The coupled SQUID qubits are a multi-level
system. The eigenenergies $E_{n}$ and eigenstates $|n)$ are obtained by
numerically solving the eigenvalue equation of $H(x_{1},x_{2})$ using the
two-dimensional Fourier-grid Hamiltonian method \cite{Chu1990}. When $x_{e1}$
and $x_{e2}\sim 0.5$ the coupled SQUID qubits have four wells in the
potential energy surface \cite{zhou-asc2004}. The four computational states
are chosen to be the lowest eigenstate of each well.

To realize single-qubit and two-qubit gates in the coupled SQUID qubits, we
apply microwave pulses\ $x_{C}$ and $x_{T}$ to $\mathcal{C}q$ and $\mathcal{T%
}q$, respectively. The interaction of the coupled qubits and pulses is $%
V\left( x_{1},x_{2},t\right) =d_{1}\left( x_{1}-x_{e1}\right) +d_{2}\left(
x_{2}-x_{e2}\right) +d_{12}$ with $d_{1}=m\omega _{LC}^{2}\left(
x_{C}+\kappa x_{T}/2\right) $, $d_{2}=m\omega _{LC}^{2}\left( x_{T}+\kappa
x_{C}/2\right) $, and $d_{12}=m\omega _{LC}^{2}\left(
x_{C}^{2}+x_{T}^{2}+\kappa x_{C}x_{T}\right) /2$. The time-dependent wave
functions of the coupled SQUID qubits, $\psi (x_{1},x_{2},t)$, are governed
by the time-dependent Schr\"{o}dinger equation $i\hbar \partial \psi
/\partial t=\left[ H(x_{1},x_{2})+V(x_{1},x_{2},t)\right] \psi $. To solve
this equation we expand $\psi $ in terms of the first 20 eigenstates of the
coupled qubits: $\psi =\sum_{n}c_{n}\left( t\right) |n)$. The expansion
coefficients $c_{n}(\tau )$ are calculated by numerically solving the matrix
equation $i\partial c_{n}\left( \tau \right) /\partial \tau =\sum_{n^{\prime
}}H_{nn^{\prime }}^{R}\left( \tau \right) c_{n^{\prime }}\left( \tau \right) 
$ using the split-operator method \cite{Hermann1988}, where $\tau =\omega
_{LC}t$ is the dimensionless time and $H_{nn^{\prime }}^{R}=\left[
E_{n}\delta _{nn^{\prime }}+\left( n\left\vert V\right\vert n^{\prime
}\right) \right] /\hbar \omega _{LC}$ is the reduced Hamiltonian matrix
element. The probability of being in the state $|n)$ is thus $\left\vert
c_{n}(\tau )\right\vert ^{2}$. For single-qubit gates two resonant pulses, $%
x_{C1}=x_{C10}\cos \left( \omega _{C1}t\right) $ with $\omega _{C1}=\Delta
E_{13}/\hbar $ and $x_{C2}=x_{C20}\cos \left( \omega _{C2}t\right) $ with $%
\omega _{C2}=\Delta E_{24}/\hbar $, are applied simultaneously to the $%
\mathcal{C}q$ so that $x_{C}=x_{C1}+x_{C2}$. For controlled two-qubit gates
such as the CNOT gate one resonant pulse $x_{T}=x_{T0}\cos \left( \omega
_{T}t\right) $ with $\omega _{T}=\Delta E_{34}/\hbar $ is applied to the $%
\mathcal{T}q$. To minimize the possible intrinsic gate errors caused by
leakage to unintended states, we select the parameters of the coupled SQUID
qubits using the independent transition approach \cite{Zhou05}. For SQUIDs
with $L=100$ pH, $C=40$ fF, and $\beta _{L}=1.2$ one set of the better
working parameters is $x_{e1}=0.499$, $x_{e2}=0.4998$, and $\kappa =5\times
10^{-4}$, which will be used in the calculation.

\begin{figure}[ptb]
\begin{center}
\psfig{file=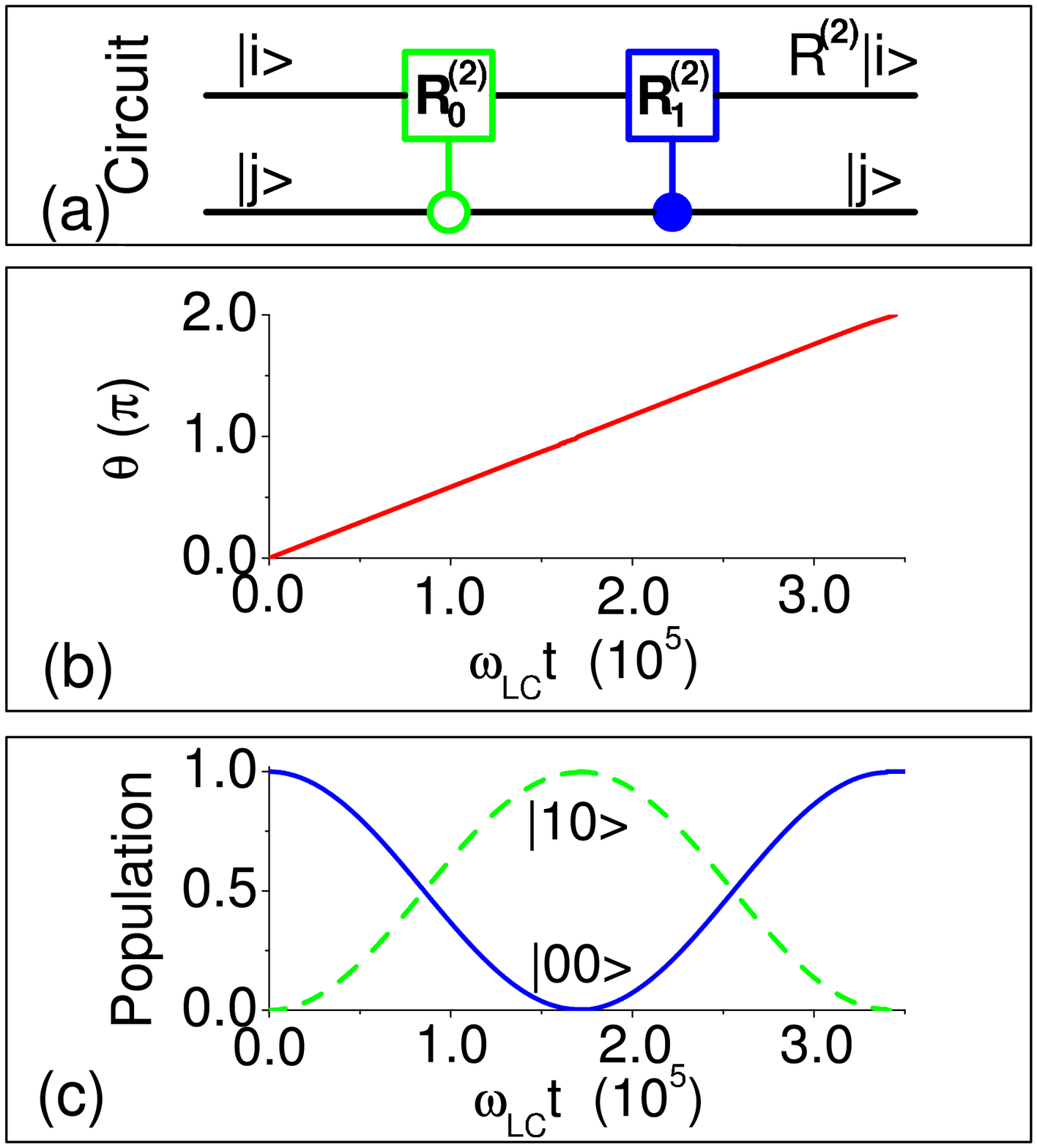,height=6cm,width=5cm,angle=90} 
\end{center}
\caption{Arbitrary single-qubit rotation about an axis perpendicular to $z$
axis in the coupled SQUID flux qubits. (a) Quantum circuit: the state $%
\left\vert ij\right\rangle $ evolves to the state $R^{\left( 2\right) }(%
\protect\theta )\left\vert ij\right\rangle =\left[ R^{\left( 2\right) }(%
\protect\theta )\left\vert i\right\rangle \right] \left\vert j\right\rangle $
after the two conditional two-qubit rotations. (b) The rotation angle $%
\protect\theta $ vs. pulse width. (c) Populations of the states $\left\vert
00\right\rangle $ and $\left\vert 10\right\rangle $ vs. pulse width.}
\end{figure}

Most of the single-qubit gates, such as the NOT gate and Hadamard gate, can
be described by rotations of the state vector on the Bloch sphere \cite%
{Nielsen2000}. Thus we demonstrate how to realize an arbitrary single-qubit
rotation of an angle $\theta $ about an axis perpendicular to the $z$ axis, $%
R^{\left( 2\right) }(\theta )$, in the coupled SQUID qubits. Based on Eq.(%
\ref{x2}), the rotation $R^{\left( 2\right) }(\theta )$ on $\mathcal{C}q$ is
accomplished via two conditional two-qubit rotations $R_{0}^{\left( 2\right)
}(\theta )$ and $R_{1}^{\left( 2\right) }\left( \theta \right) $ as shown in
FIG. 2(a). They are realized by applying two microwave pulses $x_{C1}$ and $%
x_{C2}$ to $\mathcal{C}q$ simultaneously. The amplitudes and frequencies of $%
x_{C1}$ and $x_{C2}$ are $x_{C10}=5\times 10^{-5}$, $\omega
_{C1}=0.239\omega _{LC}=2\pi \times 19.0$ GHz, $x_{C20}=5.14\times 10^{-5}$,
and $\omega _{C2}=0.259\omega _{LC}=2\pi \times 20.6$ GHz. In Fig. 2 (b) and
(c), we plot the rotation angle $\theta $ and the populations of the states $%
\left\vert 00\right\rangle $ and $\left\vert 10\right\rangle $ (populations
of the other states remain essentially at zero) as a function of pulse width
when the initial state of the coupled qubits is $\left\vert 00\right\rangle $%
. It is shown in Fig. 2 (b) that the rotation angle $\theta $ is essentially
a linear function of pulse width. This indicates that the state of $\mathcal{%
C}q$ undergoes Rabi oscillations for which the phase angle $\Omega \tau $ is
a linear function of pulse width, where $\Omega $ is the Rabi frequency. It
is also shown in Fig. 2 (c) that from the initial state $\left\vert
00\right\rangle $ the coupled qubits evolve into the state $\left(
\left\vert 00\right\rangle +\left\vert 10\right\rangle \right) /\sqrt{2}$
after the $\pi /2$ pulses and into the state $\left\vert 10\right\rangle $
after the $\pi $ pulses. We have also computed the rotation angles and
populations for the coupled qubits with the initial states $\left\vert
01\right\rangle $, $\left\vert 10\right\rangle $,$\ $and $\left\vert
11\right\rangle $ using the same pulse sequence. In each case, the state is
transformed from $\left\vert ij\right\rangle $ to $R^{\left( 2\right)
}(\theta )\left\vert ij\right\rangle =\left[ R^{\left( 2\right) }(\theta
)\left\vert i\right\rangle \right] \left\vert j\right\rangle $ for $i,j=0,1$
with $\theta =\Omega \tau $ as expected.

\begin{figure}[ptb]
\begin{center}
\psfig{file=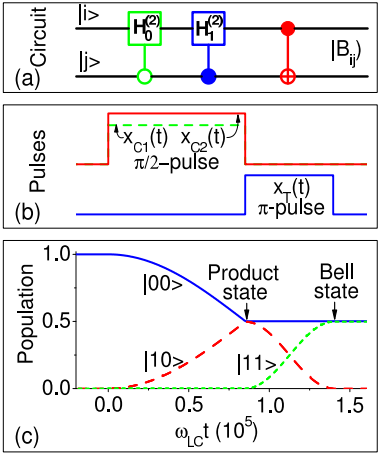,height=6cm,width=5cm,angle=90} 
\caption{Creation of the Bell states in the coupled SQUID flux qubits. (a)
Quantum circuit: the state $\left\vert ij\right\rangle $ evolves into an
entangled state $\left\vert B_{ij}\right) $ after two conditional two-qubit
Hadamard gates and one CNOT gate. (b) Microwave pulses: two $\protect\pi /2$
pulses $x_{C1}$ and $x_{C2}$ are applied to the $\mathcal{C}q$ first and
then a $\protect\pi $ pulse $x_{T}$ is applied to the $\mathcal{T}q$. (c)
Population evolution: the coupled qubits evolve into a product state $\left(
\left\vert 00\right\rangle +\left\vert 10\right\rangle \right) /\protect%
\sqrt{2}$ from the initial state $\left\vert 00\right\rangle $ after the
first two $\protect\pi /2$ pulses and then into a Bell state $\left(
\left\vert 00\right\rangle +\left\vert 11\right\rangle \right) /\protect%
\sqrt{2}$ from the product state after the second $\protect\pi $ pulse.}
\end{center}
\end{figure}

Entanglement is one of the most profound characteristics of quantum systems
which plays a crucial role in quantum information processing and
communication \cite{Nielsen2000}. The maximally entangled two-qubit states
are referred to as the Bell states. To create the Bell states from a state $%
\left\vert ij\right\rangle $, a Hadamard gate on $\mathcal{C}q$, $H^{(2)}$,
which is decomposed into two conditional two-qubit Hadamard gates $%
H_{0}^{(2)}$ and $H_{1}^{(2)}$, and a CNOT gate are commonly used [see FIG.
3(a)]. The two conditional two-qubit Hadamard gates are implemented by
applying two $\pi /2$ pulses $x_{C1}$ and $x_{C2}$ to $\mathcal{C}q$ and the
following CNOT gate is implemented by applying a $\pi $ pulse $x_{T}$ to $%
\mathcal{T}q$, as shown in FIG. 3(b). The amplitudes and frequencies of $%
x_{C1}$ and $x_{C2}$ are the same as those used in FIG. 2 and those of $%
x_{T} $ are $x_{T0}=5\times 10^{-5}$ and $\omega _{T}=0.0592\omega
_{LC}=2\pi \times 4.7$ GHz. In FIG. 3(c), we plot the population evolution
of the computational states when the initial state is $\left\vert
00\right\rangle $. It is shown clearly that the coupled qubits evolve first
into a product state $\left( \left\vert 00\right\rangle +\left\vert
10\right\rangle \right) /\sqrt{2}$ from the initial state $\left\vert
00\right\rangle $ after the $\pi /2$ pulses and then into a Bell state $%
\left( \left\vert 00\right\rangle +\left\vert 11\right\rangle \right) /\sqrt{%
2}$ after the subsequent $\pi $ pulse. We have also calculated the
population evolution for the coupled qubits being initially in $\left\vert
01\right\rangle $, $\left\vert 10\right\rangle $, and $\left\vert
11\right\rangle $, respectively. The final state in each case is also one of
the expected Bell states.

In summary, we showed that in a coupled two-qubit system \textit{any
single-qubit gate can be realized via two conditional two-qubit gates} and
that \textit{any conditional two-qubit gate can be implemented with a
manipulation analogous to that used for a controlled two-qubit gate}. Based
on this universal property of single-qubit gates we present a general
approach to implement the universal single-qubit and two-qubit gates in the
same coupled two-qubit system with fixed always-on coupling. This approach
is demonstrated by using a unit of two SQUID flux qubits with realistic
device parameters and constant always-on coupling. Compared to other methods
our approach has the following characteristics and advantages: (1) The
approach is universal as long as each qubit can be locally addressed; (2) No
additional decoherence from the hardware added to control inter-qubit
coupling; (3) Gate error induced by the population propagation from one
qubit to another is completely eliminated; (4) The architecture for both
hardware (circuits) and software (pulse sequence) is much simplified. This
approach can be readily extended to multi-qubit systems and other types of
solid-state qubit systems in which each qubit can be individually addressed.
Therefore, it is very promising for realizing universal gates with minimum
resource and complexity and maximum efficiency.

This work is supported in part by the NSF (DMR-0325551) and by AFOSR, NSA,
and ARDA through DURINT grant (F49620-01-1-0439).

\bibliographystyle{apsrev}
\bibliography{t-squid1}

\end{document}